\documentclass[10pt]{iopart}
\usepackage{iopams} 
\usepackage{setstack} 
\usepackage{graphicx}

\newcommand{\eqref}[1]{(\ref{#1})}

\newcommand{\ket}[1]{\vert{#1}\rangle}
\newcommand{\bea}{\begin{eqnarray}}
\newcommand{\eea}{\end{eqnarray}}
\newcommand{\beq}{\begin{equation}}
\newcommand{\eeq}{\end{equation}}

\begin{document}

\title[Effective Hamiltonian in multidimensional active spaces]{Global integration of the Schr\"odinger equation within
the wave operator formalism:
The role of the effective Hamiltonian in multidimensional active spaces}

\author{Georges Jolicard\dag, Arnaud Leclerc\S, David Viennot\dag, and John P. Killingbeck\ddag}
\address{\dag\ Institut Utinam, UMR CNRS 6213, Universit\'{e} de Franche-Comt\'{e}, Observatoire de Besan{\c c}on,
41 bis avenue de l'Observatoire, BP1615, 25010 Besan{\c c}on Cedex, France}
\address{\S\ SRSMC, UMR CNRS 7565, Universit\'e de Lorraine, 1 Bd Arago 57070 Metz, France}
\address{\ddag School of Engineering, Robert Blackburn Building, University of Hull, Hull HU6 7RX, UK}

\eads{\mailto{georges.jolicard@utinam.cnrs.fr}, \mailto{arnaud.leclerc@univ-lorraine.fr},
\mailto{david.viennot@utinam.cnrs.fr}, \mailto{j.p.killingbeck@hull.ac.uk}}

\begin{abstract}

A global solution of the Schr\"odinger equation, obtained recently
within the wave operator formalism for explicitly time-dependent Hamiltonians
[J. Phys. A: Math. Theor. 48, 225205 (2015)], 
is generalized to take into account the case of multidimensional active spaces. 
An iterative algorithm is derived to obtain the Fourier series of
the evolution operator issuing from a given multidimensional active subspace and then the
effective Hamiltonian corresponding to the model space is computed 
and analysed as a measure of the cyclic character of the dynamics.
Studies of the laser controlled dynamics of diatomic models
clearly show that a multidimensional active space is required if
the wavefunction escapes too far from the initial subspace. 
A suitable choice of the multidimensional active space,
including the initial and target states, increases the cyclic character and 
avoids divergences occuring when one-dimensional active
spaces are used. 
The method is also proven to be efficient in describing dissipative processes such as photodissociation.

\end{abstract} 




\maketitle

\section{ Introduction}

There are several numerical approaches for dealing with problems of quantum dynamics 
if a complicated time dependence is present in the Hamiltonian.  
Here we are particularly concerned with problems in which a molecule is subjected to an external field. 
In such cases, fast oscillations of the (classical) electromagnetic field and of 
the wavefunction must be taken into account accurately, as for example in
quantum control algorithms which involve 
the intensive use of numerical wavefunction propagation to handle strong molecules-laser coupling 
\cite{shapiro2003,werschnik2007}. 
If the external field is a continuous wave with a constant envelope and a single well defined frequency, 
then the Floquet formalism is very well adapted and gives a basis of periodic solutions to the Schr\"odinger equation (the Floquet states) 
\cite{Shirley,Sambe}. 
In this case the problem is simplified to require only 
the study of a single optical period of the field. 
The Floquet states are generally not directly calculated in the time domain but rather via their Fourier coefficients in the frequency domain. 
It shall also be noted that the Floquet approach 
can be used to prove the equivalence of the purely
quantum approach and the semi-classical one (under some precise conditions) in the strong field
regime \cite{Bialynicki,Guerin2003}. 
The standard Floquet approach must be generalized if the field is not a continuous wave. 
If the intensity or the frequency of the electromagnetic field are slowly varied, good approximate results can still be obtained from the adiabatic variant of the Floquet
 theory \cite{Guerin2003}. 
However, the only exact and rigorous way to determine the wave packet evolution in the general case of a rapidly varying or chirped pulse
 is to use an exact wavepacket propagation. 
This is the topic treated here. Floquet states can still be calculated but they are called
generalized Floquet states and are associated with the total duration of the interaction. 

We develop a propagation method which makes use of two important ideas. 
The first idea is that, even in the case of a non-adiabatic or complicated pulse, 
time-dependence can still be described by using a Fourier basis set, 
instead of using the differential step by step scheme adopted by some well established methods \cite{leforestier1991,balakrishnan1997}. 
Fourier grids methods have long been used to describe the molecular coordinates in wavepacket propagation
\cite{kosloff1983}.  
They have not been so widely used to treat the time dependence of wavepackets, because the real wavepacket is not generally time-periodic.
 We shall see that it is possible to resolve this apparent inconsistency by using  absorbing potentials such as those introduced in ref. \cite{jolicard2004,leclerc2011}. 
The second important idea which we use is that the quantum states are not all equally significant in a given propagation. 
For example in a quantum control problem, the initial and the target states are of particular interest while
in a STIRAP process three states are mostly concerned \cite{vitanov2001}, etc. 
It is clearly appropriate to reflect this hierarchy of importance in the numerical approach used to solve the Schr\"odinger equation. 
In this article we take advantage of the subspace decomposition idea by relying on the time-dependent wave operator theory \cite{Jol3}.

The two main ideas outlined above  have been combined to propose a global integration method for the Schr\"odinger equation 
within the wave operator formalism in ref. \cite{Lecl2015}.
This first formulation was intended only for hermitian 
Hamiltonians and was limited to the use of one-dimensional active spaces; it
appears to be
efficient for investigating near-adiabatic
evolutions. However for many processes such as dissociations, 
ionization and more general strong non-adiabatic interactions 
this first version cannot be used. 
It is thus necessary to generalize it: first, to be able to describe dissipative processes 
driven by non-hermitian Hamiltonians and second (and most importantly),
to handle multidimensional active spaces for strong non-adiabatic evolutions, relevant in situations where
the wavefunction escapes too far from a one-dimensional model space.

In section \ref{iterative_proc} 
this generalization is achieved by introducing small multidimensional active spaces
in place of the one-dimensional ones 
and by using  asymptotic complex absorbing potentials to discretize correctly the interacting molecular continua. 
Iterative formulae 
are derived by combining the two techniques used in 
the first version of the algorithm : 
the Time Dependent Wave Operator (TDWO) formalism and the calculation of time-dependent integrals 
by using fast Fourier techniques (FFT). 
This leads to an iterative solution of the global dynamical problem. 
Low dimensional effective Hamiltonians are analysed as tools 
for measuring the more or less cyclic character of the dynamical processes 
and close relationships 
are revealed between this cyclic character and the relevance of using the TDWO theory. 
In section \ref{num_examples} 
we illustrate the algorithm 
and the theoretical discussion 
by studying a complete vibrational population transfer between the two
wells of a model energy surface representing a system under the influence of laser fields. 
A second example describes 
the dissociative dynamics of the H$_2^+$ molecular ion
coupled to a laser field. This example confirms the ability 
of the method to treat dissipative processes, 
if an appropriate multidimensional active space is used. 
Section \ref{conclusion} gives a discussion and some concluding remarks.

\section{Time dependent wave operator using multidimensional model spaces \label{iterative_proc}}%

\subsection{Iterative calculation of the wave operator \label{itcalc}} 

Let $\mathcal{H}$ be the Hilbert space associated with a molecular system and let $S_o$ be a model  subspace of dimension $m$
which includes the initial molecular state. 
The orthogonal projector corresponding to the model space is called $P_o$, with 
$P_o^2=P_o$, $P_o^{\dag}=P_o$, $tr(P_o)=m$. 
The time-dependent wave operator associated with the two subspaces $S_o$ and $S(t)$ is defined as \cite{Jol3}:
%
%
\begin{equation}
\Omega (t)=P(t)(P_oP(t)P_o)^{-1}=U(t,0;H)(P_oU(t,0;H)P_o)^{-1}
\label{WOD}
\end{equation}
where $P(t)$ are the projectors of the successive model spaces $S(t)$, with $S(t=0)=S_o$ and
where $U$ represents the quantum evolution operator associated with the Hamiltonian $H(t)$. 
$P(t)$ is the time-dependent transformation of $P_o$ under the influence of the Schr\"odinger equation,
i.e. $P(t)=U(t,0;H)P_o U^{\dagger}(t,0;H)$. 
The time-dependent wave operator is a generalization of the M{\o}ller wave operator 
and can be used to deduce the true dynamics from the dynamics 
within the $m-$dimensional subspace, 
governed by $H^{eff}=P_o H\Omega$: 
%
\beq
U(t,0;H)P_o=\Omega (t) U(t,0;H_{eff}).
\label{evop}
\eeq
However, the wave operator exists only if the
Fubini-Study distance between $P_o$ and $P(t)$ is small:
 $dist_{FS}(P_o,P(t)) < \frac{\pi}{2}$ \cite{Vien1}.
This means that the real dynamics should not make the wavefunction  escape too far from the selected initial subspace. 
For strong couplings the limit value $\frac{\pi}{2}$ can easily be reached
if the subspace is the one-dimensional subspace associated with the initial state \cite{Lecl2015}. 
The choice of a multidimensional model space $S_o$
which includes all the states which interact strongly with the inital state
can then reduce the Fubini-Study distance and 
make the calculation possible. 

The wave operator satisfies a Bloch equation in the extended Hilbert space $\mathcal{H} \otimes L_o^2([0,T])$
($T$ is the total duration of the dynamical process and $L_o^2([0,T])$ denotes the space of square integrable functions on
$[0,T]$ with periodic boundary conditions),
%
%
\begin{equation}
H_F(t)\Omega(t)=\Omega(t)(H(t)\Omega(t))=\Omega(t) H_{eff}(t).
\label{BAEQ}
\end{equation}
In the above equation $H_F(t)$ is the Floquet hamiltonian associated with the total duration,
$H_F(t)=H(t)-i\hbar \partial/\partial t$. 
The main difficulty arising in this formalism is the integration of equation (\ref{BAEQ}). 
Although the structure of eq. (\ref{BAEQ}) looks simple, this simplicity is deceptive.
First, this equation is not a pure series of instantaneous eigen-equations, since the time-derivative 
present in $H_F$ in the left hand side couples together 
the values of $\Omega(t)$ at different times. 
Second, the integration of the 
Schr\"odinger equation within this formalism 
generally requires us to enforce the initial conditions. 
If the interaction is located on the finite time interval $[0,T]$ 
and if the initial wave function $\Psi (t=0)$ is strictly included within the subspace $S_o$, then the integration of equation (\ref{BAEQ}) 
using tools such as Fourier basis sets
and Fast Fourier Transform (FFT)
can only give strictly T-periodic solutions. 
Such solutions do not generally satisfy 
the correct initial conditions except under special 
adiabatic circumstances (for example 
$\Psi (t=T) \propto \Psi (t=0)$). 
To recover the correct initial conditions by using a periodic basis set, 
we assume that the physical interaction is restricted to a time interval $[0,T_0]$, which is 
shorter than the total time interval $[0,T]$ used to describe it numerically, with $T_0<T$. 
Then a time-dependent absorbing potential is introduced over the artificial 
time extension $[T_o,T]$. 
$Q_o$ being the projector on the space complementary to the active space $S_o$ ($P_o+Q_o=1$),
the suitable absorbing potential is \cite{jolicard2004,leclerc2011}
%
%
\begin{equation}
V_{abs}=-iV_{opt}(t) \; Q_o
\label{VOPT}
\end{equation}
where $V_{opt}(t)$ is a real positive function localised on the time interval $[T_o,T]$. The results of the
dynamics are analysed at the final physical time $T_o$, 
the behaviour of $\Omega$ 
during the asymptotic time extension being purely artificial
and having no influence back on the physical interval $[0,T_0]$. 

Before going into more details about the numerical algorithm used to solve eq. (\ref{BAEQ}),
it is useful to clarify the general framework of the calculation. 
In the following the Hamiltonian $H(t)$ which drives the dynamics
is the sum of an unperturbed Hamiltonian $H_o$, a time-dependent 
coupling term $V(t)$ 
(corresponding to the electric dipole coupling $-\vec{\mu}.\vec{E}(t)$ in laser-molecule experiments) 
and a time-dependent absorbing potential $V_{abs}$ (eq.\ref{VOPT}). 
The Hilbert space $\mathcal{H}$ is assumed to be 
truncated to a finite-dimensional space. 
If the potential energy curves in $H_o$ are dissociative, a radial optical potential $\tilde{V}_{opt}(r)$ 
is introduced to discretise the continuum associated with the dissociative radial coordinate.
Finally the Hamiltonian operator is ($r$ is a composite molecular coordinate)
%
%
\beq
H(r,t) = H_o(r) + V(r,t) + \tilde{V}_{opt}(r) + V_{abs}(t).
\label{hamiltonien}
\eeq
A complete zeroth-order basis set $\{|j\rangle\}_{j=1,N_m}$ 
is made up of $N_m$ eigenvectors of $H_o$ (or $H_o+\tilde{V}_{opt}(r)$ if needed). 
A Fourier basis set $\{|n \rangle\}_{n=1,N_t}$ with $\langle t|n\rangle=\exp (i 2\pi n t/T)$
is used to represent the space $L_o^2[0,T]$. Finally, if the active space is of dimension $m$, 
the wave operator $\Omega$ is represented by $N_t$ rectangular matrices whose size is $(N_m\times m)$ corresponding
to each of the $N_t$ values of the Fourier frequencies (or equivalently to the $N_t$ discrete sampling values in time).
This series of matrices can also be recast into one single $((N_m.N_t) \times m)$ rectangular matrix. 

By projecting equation (\ref{BAEQ}) on the left into the complementary space with projector
$Q_o$, with $P_o + Q_o =1$, and by introducing the reduced wave operator $X=Q_oXP_o$, with $\Omega= P_o +X$, 
a new reduced equation is obtained:
%
%
\begin{equation}
Q_o(1-X(t))H_F(t)(1+X(t))P_o=0
\label{RBAEQ}
\end{equation}
where $H_F(t)$ includes the absorbing potential $V_{abs}(t)$.
An iterative solution of equation (\ref{RBAEQ}) has been proposed in \cite{Lecl2015} for a one-dimensional active space. 
We now derive the solution in the case of a multidimensional active space. 
By assuming that eq. (\ref{RBAEQ})
is not perfectly satisfied at the finite iteration order $(n)$, the right hand side of this equation being equal
to $\Delta^{(n)}(t)=Q_o(1-X^{(n)}(t))H_F(t)(1+X^{(n)}(t))P_o$ 
instead of
zero, one can introduce the increment $\delta X^{(n)}(t)$ such that $X^{(n+1)}(t) = X^{(n)}(t) + \delta X^{(n)}(t)$ exactly solves the equation. 
Expanding eq. (\ref{RBAEQ}) leads to: 
%
%
\begin{equation}
i \hbar \frac{\partial}{\partial t}\delta X^{(n)}(t) = \Delta^{(n)}(t)-\delta X^{(n)}(t) H_{eff}^{(n)}(t)
+ \tilde{H}^{(n)}_{diag}(t) \; \delta X^{(n)}(t)
\label{1diff}
\end{equation} 
with
%
%
\begin{eqnarray}
\left \lbrace \begin{array}{l}
\tilde{H}^{(n)}_{diag}(t)=Q_o[H(t)-X^{(n)}(t)H(t)]_{diag} Q_o \\
H_{eff}^{(n)}(t)=P_o H(t) \Omega^{(n)}(t)=P_o H(t)(P_o+X^{(n)}(t))
\end{array} \right .
\label{TOTAL}
\end{eqnarray}
To derive eq. (\ref{1diff}), some approximations, previously tested in ref. \cite{Lecl2015} 
with $m=1$, 
have been introduced.
The quadratic terms with respect to the increment $(\delta X^{(n)}(t))$  and the
non-diagonal elements of  $Q_o[H(t)-X^{(n)}(t)H(t)]Q_o$ 
have been neglected. This leads in the multidimensional case to the following rigorous solution of equation (\ref{1diff}),
%
%
\begin{eqnarray}
&\delta X^{(n)}(t) &= U(t,0;\tilde{H}^{(n)}_{diag})  \nonumber \\
&&\times \left[ \frac{1}{i\hbar}\int_0^t U^{-1}(t',0;\tilde{H}^{(n)}_{diag})
\Delta^{(n)}(t') U(t',0;H_{eff}^{(n)})dt' \right] \nonumber \\
&& \times U^{-1}(t,0;H_{eff}^{(n)}),
\label{Sol1}
\end{eqnarray}
where the letters 
$U$ represent the quantum evolution operators. 
At a fixed time $t$, 
$\Delta^{(n)}(t)$ is a $(N_m \times m)$ rectangular matrix,
$U(t,0;\tilde{H}^{(n)}_{diag})$ is a diagonal matrix with $(N_m)$ entries 
and $U(t,0;H_{eff}^{(n)})$ is a small $(m\times m)$ matrix.

\subsection{Discrete implementation of the iterative solution}

To integrate eq. (\ref{Sol1}), a procedure based on 
Fast Fourier Transforms is used. 
A discrete finite time-grid is introduced 
on the time interval $[0,T]$: 
%
%
\begin{equation}
t_j=\frac{jT}{N_t}, \;\; j=0,\ldots,N_t-1, 
\label{TGRID}
\end{equation}
together with the corresponding frequency representation: 
%
%
\begin{eqnarray}
\left \lbrace \begin{array}{l}
\nu_j=\frac{j}{T}, \;\;j=0,\ldots,\frac{N_t}{2}-1, \\
\nu_{N_t/2}=-\frac{N_t}{2T},\\
\nu_j=-\nu_{N_t-j}, \;\; j=\frac{N_t}{2}+1,\ldots,N_t-1
\end{array} \right .
\label{NUGRID}
\end{eqnarray}
In eq.(\ref{Sol1}), the various terms are discretized by using the time-grid representation (eq. (\ref{TGRID})).
The calculation of $U(t_j,0;H^{(n)}_{eff})$ takes advantage of the small dimension of the $(m \times m)$ matrix $H^{(n)}_{eff}$. 
The evolution operator associated with $H^{(n)}_{eff}$ is calculated using
%
%
\begin{equation}
U(t_k,0;H^{(n)}_{eff})=\prod_{j=1}^{k}U(t_j,t_{j-1};H^{(n)}_{eff}),\;k=0,\ldots , N_t-1.
\label{Produc}
\end{equation}
The time evolution associated with $H^{(n)}_{eff}$ between two adjacent discrete time is 
approximated by
%
%
\begin{equation}
U(t_{j},t_{j-1};H^{(n)}_{eff}) = \exp \left(-\frac{i}{\hbar}\int_{t_{j-1}}^{t_{j}} H^{(n)}_{eff}(t') dt' \right)
\label{Ueff}
\end{equation}
and its action on an arbitrary vector is obtained by diagonalizing the matrix 
$\int_{t_{j-1}}^{t_{j}} H^{(n)}_{eff}(t') dt'$. 
In eq. (\ref{Sol1}) the matrix $H^{(n)}_{eff}$ exhibits
large time variations and it is essential to retain the exact expression and to make a precise calculation of
$U(t',0;H^{(n)}_{eff})$. 
To do this, numerous numerical integrals are needed in eq. (\ref{Ueff}). 
At each iteration order $(n)$ the $m^2$ components $[H^{(n)}_{eff}]_{k,l}$ are integrated on the time interval
 $[0,T]$ by using the FFT procedure
proposed in ref.\cite{Lecl2015}. 
This method requires only two FFT to obtain the $N_t$ definite integrals corresponding to 
all the intermediate intervals $t_j$ in eq. (\ref{Ueff}). 

The calculation of the evolution operator associated with the matrix $H^{(n)}_{diag}$
can also be done by using a Fourier algorithm. 
This matrix includes the asymptotic absorbing potential $V_{opt}(t)$ 
which can produce numerical instabilities during the discrete calculation of the integrals
(because of its real exponential behavior). 
To solve this difficulty, this absorbing potential has been neglected in the term 
$U^{-1}(t',0;\tilde{H}^{(n)}_{diag})$ 
inside the integral $\int_0^t (\ldots)$ (c.f. eq.(\ref {Sol1}))
and has been simultaneously preserved in the term $U(t,0;\tilde{H}^{(n)}_{diag})$ on the left 
in order to impose the correct initial conditions. 
These apparently arbitrary approximations are justified a posteriori 
by the convergence of the iterative solution. 
The matrix $\tilde{H}^{(n)}_{diag}$ is diagonal. 
The operator $U^{-1}(t',0;\tilde{H}^{(n)}_{diag})$ can be easily calculated as
%
%
\begin{equation}
U^{-1}(t',0;\tilde{H}^{(n)}_{diag})=\exp \left( \frac{i}{\hbar}H_o t' \right) \exp \left( \frac{i}{\hbar}\int^{t'}_0\delta \tilde{H}^{(n)}_{diag}dt'' \right). 
\label{Udiag}
\end{equation}
with $\delta\tilde{H}^{(n)}_{diag}=\tilde{H}^{(n)}_{diag}-H_o$.

Introducing (\ref{Udiag}) in eq.(\ref{Sol1}) and using the new time-dependent matrix $\Lambda^{(n)}(t)$ 
defined as
%
%
\beq
\Lambda^{(n)}(t)=\exp \left(\frac{i}{\hbar}\int^{t}_0 \delta{\tilde{H}}^{(n)}_{diag} dt' \right) \,  \Delta^{(n)}(t) \, U(t,0;H^{(n)}_{eff}),
\label{DeltTTF}
\eeq
eq. (\ref{Sol1}) can be rewritten as
%
%
\bea
\delta X^{(n)} &=& e^{-\frac{i}{\hbar} H_o t} e^{-\frac{i}{\hbar} \int_0^t \delta \tilde{H}_{diag}^{(n)} dt'}
\left( \frac{1}{i\hbar} \int_0^t e^{\frac{i}{\hbar} H_o t'} \Lambda^{(n)} (t') dt' \right) \nonumber \\
 && \times U^{-1} (t,0;H^{(n)}_{eff}).
\label{deltaXint}
\eea
Using the Fourier transform  $\tilde{\Lambda}$ of $\Lambda$, such as
%
%
\beq
{\Lambda}^{(n)}( t ) = \int_{\nu} \tilde{\Lambda}^{(n)}(\nu) \exp( i 2\pi\nu t) d\nu,
\eeq
eq. (\ref{deltaXint}) can be rearranged in the following form 
%
%
%
%
%
%
\begin{eqnarray}
\delta X^{(n)}(t)&=&\exp \left(-\frac{i}{\hbar}\int^{t}_0\delta \tilde{H}^{(n)}_{diag}dt' \right) \nonumber \\
&\times& \left[-Z^{(n)}(t)+\exp \left(-\frac{i}{\hbar}H_o t \right) Z^{(n)}(t=0) \right]
\nonumber \\
&\times & U^{-1} (t,0;H^{(n)}_{eff})
\label{Eqfinal}
\end{eqnarray}
where the matrix $Z^{(n)}$ 
is obtained by using two back and forth Fourier transforms, 
%
\begin{equation}
Z^{(n)}(t)=FT^{-1}_{(t)} \left[ \frac{FT_{(\nu)}(\Lambda^{(n)}(t) )}{H_o +2\pi \hbar \nu} \right].
\label{Zeq}
\end{equation}
The matrix elements of $Z^{(n)}$ in eq. (\ref{Zeq}) are calculated by using the discretized time/frequency introduced in
eqs (\ref{TGRID}) and (\ref{NUGRID}) and by approximating the Fourier Transforms by discrete Fast Fourier Transforms.
Finally the incrementation rule
%
%
\begin{equation}
X^{(n)}=X^{(n-1)} +\delta X^{(n-1)}
\label{Incr}
\end{equation}
and the equations (\ref{DeltTTF}), (\ref{Eqfinal}) and (\ref{Zeq})
constitutes 
the iterative scheme leading to the wave operator and 
to all the columns of the evolution operator (eq. (\ref{evop})) issuing from subspace $S_o$. 
Any wavefunction whose initial state belongs to $S_o$ 
can be written as
%
%
\begin{equation}
|\Psi^{(n)}(t)\rangle=(P_o + X^{(n)}(t)) U(t,0;H^{(n)}_{eff})|\Psi (t=0)\rangle.
\label{PropGlob}
\end{equation}
Some approximations have been introduced 
within the above iterative scheme. 
Nevertheless the algorithm is global 
and these approximations 
cannot induce any of the cumulative errors found in a standard differential propagation scheme. 
The iterative procedure is stopped when the
following convergence criteria is satisfied:
%
%
\beq
||\delta X^{(n)}||^2/||X^{(n)}||^2 \leq \epsilon \\
\label{IterStop}
\eeq
where $\epsilon$ is a fixed convergence factor
and $||.||$ denotes the Frobenius norm. 
The solution which satisfies 
eq. (\ref{IterStop}) is necessarily the correct and unique solution of the
propagation problem with  
an accuracy specified by $\epsilon$. 

We would like to stress the fact that a calculation with a subspace of 
dimension $m'>m$ gives more results than the calculation
with a subspace of dimension $m$, because more columns of the evolution operator are obtained. 
At the same time more columns have to be treated by the Fourier transform steps. 
As long as $m$ remains small this part of the calculation takes most of the CPU time. 
This CPU time increase is linear with $m$ and even if other parts of the calculation 
are not (several matrix products cost $m^2$ and the effective Hamiltonian diagonalizations cost $m^3$) 
the final CPU time increase is about linear. Moreover an increase of $m$ can
produce a strong acceleration of the convergence, leading to a final decrase of the total CPU time.

\subsection{Cyclic dynamics and wave operator}

The iterative solution proposed in the previous subsection is relevant for dynamical processes in which the wavefunctions does not escape very far from the model subspace. 
The quantum distance between the dynamically developing subspace and the model space $S_o$ can be monitored by using the Fubini-Study distance between
$S_o$ and $S(t)$, namely $dist_{FS}(S_o,S(t))$. 
This distance goes from $0$, when the two subspace are equal, and increases to reach its maximum possible value 
of $\pi/2$ when the subspaces are orthogonal. This situation could make the iterative algorithm diverge. 
If $|i\rangle$ is the initial state and 
the wavefunction is $\ket{\Psi_i(t)} = U(t,0;H) \ket{i}$, 
the Fubini-Study distance corresponding to a one-dimensional active space is simply equal to: 
%
%
%
\begin{equation}
dist_{FS}(S_o,S(t))= 
\arccos \left( \parallel \langle i \vert \Psi_i (t) \rangle \parallel \right) = \arccos \left( \frac{1}{\parallel \Omega (t) \parallel} \right).
\label{FSD}
\end{equation}
Eq. (\ref{FSD}) indicates that the limit value is reached if the survival probability vanishes at a given time,
which is equivalent to say that the wave operator diverges at the same time. 
There are two different approaches to overcome the possible difficulty due to nearly-orthogonal subspaces. 
In ref. \cite{Vien1} time-dependent adiabatic deformations of the active space are shown to be efficient 
for describing almost adiabatic quantum systems.
Here
we adopt another point of view and we try to improve the 
integration scheme by using multidimensional active spaces in the global integration procedure. 
In the multidimensional case, the above definition of the quantum distance is generalized to give
%
%
%
%
\beq
dist_{FS}(S_o,S(t))= \arccos \vert \det{ \left( \bf{P_o^{\dagger} . ( U(t,0;H) P_o)} \right)}  \vert .
\label{FSD2}
\eeq
 where $ \bf{P_o^{\dagger} . ( U(t,0;H) P_o) }$ designates the matrix representation of the operator
$ P_o^{\dagger} . ( U(t,0;H) P_o) $ .
The conditions for this distance to reach its limit value of $\pi/2$ are less easily satisfied
and the convergence of the iterative scheme becomes more robust.

In this context cyclic dynamics are a particular case of special interest. 
In this case the wavefunction issuing from subspace $S_o$ comes back to $S_o$ at time $T$. 
A cyclic dynamics, with respect to the subspace $S_o$ of dimension $m$, 
can be defined as follows: 
$\forall i \in [1,m]$, let $\Psi_i^o(t)$ be the solution of the Schr\"odinger equation with the
initial condition $\Psi_i^o(t=0)=|i\rangle$, then there exists a unitary gauge transformation $W$ such that
$\Psi_i^o(t)=\sum_{j=1}^m W_{ji}(t) \tilde{\Psi}_j^o(t)$ 
with $\tilde{\Psi}_i^o(0)=|i\rangle=\tilde{\Psi}_i^o(T)$. 
This condition implies 
%
%
%
\beq
P_o \Psi(T)=\Psi(T)
\eeq
and by taking eq. (\ref{PropGlob}) into account,
%
%
\begin{equation}
X^{N}(t=0)=X^{N}(t=T)=0.
\label{Cycl}
\end{equation}
This means that the TDWO is also cyclic with $\Omega(T)=\Omega(0)=P_o$. 

For any Hamiltonian,
a perfect 
artificial
cyclic dynamics can always be obtained, independently of the $m$ value, by adding a time-dependent absorbing potential
(eq.(\ref{VOPT})) on the time extension $[T_o,T]$. 
This constrained cyclicity is a numerical artefact
which can be seen as the multidimensional generalization of the constrained adiabatic trajectory method
(CATM) of ref. \cite{jolicard2004,leclerc2011}. 
A true cyclic (or maybe quasi-cyclic) dynamics can also be obtained in particular cases
by the natural evolution on $[0,T_0]$, 
without introducing any artificial absorbing potential. 
In the case of a natural cyclicity, eq.(\ref{Cycl}) may not be rigorously satisfied. 
Nevertheless, the asymptotic values $X^N(0)=X^N(T)$, even when not exactly equal to zero, 
may be small enough to consider the dynamics as cyclic. We will see later
that such a naturally cyclic situation has beneficial consequences on the convergence of the iterative process. 
The main question is to select, 
with or without an absorbing potential, 
the best active space at a fixed small degeneracy, with order $m \not= 1$. 
In most cases the choice of the model space should be based on physical considerations,
thus including the initial state, strongly coupled states and states corresponding to resonant transitions. 
Choosing too small a dimension $m$ may produce too large a Fubini-Study distance between $S_o$ and $S(t)$. 
Including more states in the subspace avoids such problems. 

Cyclic dynamical processes belong to
the framework of the non-abelian, non-adiabatic Berry phase formulation,
known as the non-Abelian Aharonov-Anandan phase formulation \cite{Aha}.
This theory is consistent with the wave operator approach used in this article.
More precisely
if $\{\Psi_i^o(t)\}_i$ is a non-abelian
parallel transport associated with the section $\{\tilde{\Psi}_i^o\}_i$, then $\{\Omega \Psi_i^o\}_i$ is
a non-abelian parallel transport associated with the section $\{\Omega \tilde{\Psi}_i^o\}_i$ \cite{Vien2}.

In this cyclic context, important results 
can be derived by considering the expansion of the wave function on
the Floquet eigenstates basis set. 
In the extended Hilbert space $\mathcal{H} \otimes L_o^2([0,T])$, 
the generalized Floquet eigenstates are defined as solutions of the following equation:
%
%
\beq
H_F \ket{\lambda_{j,n}} = E_{j,n} \ket{\lambda_{j,n}} 
\eeq 
where $H_F=H-i\hbar \partial/\partial t$.
The label $n$ corresponds to the different Floquet blocks of eigenstates 
and $j$ distinguishes the states within each block
(this index is associated with the molecular Hilbert space). 
The eigenfunctions $|\lambda\rangle$ are $T-$periodic on the full interaction interval $[0,T]$ 
and satisfy the orthonormality condition \cite {Chu89}
%
%
\begin{equation}
\langle\langle \lambda_{i',n'}|\lambda_{i,n}\rangle\rangle=\frac{1}{T}\int_0^T dt \langle
\lambda_{i',n'}|\lambda_{i,n}\rangle=\delta_{i',i}\delta_{n',n}.
\label{Ortho}
\end{equation}
The wavefunction can be expanded 
on a complete set of Floquet eigenvectors, 
%
%
\begin{equation}
|\Psi (t)\rangle=\sum_{j=1}^{N_m}\sum_{n=0}^{N_t-1} e^{-iE_{{j,n}}t/\hbar}|\lambda_{j,n}(t)\rangle
\langle\langle \lambda_{j,n}|\Psi(t=0) \rangle\rangle.
\label{DevLam}
\end{equation}
By taking into account the periodicity of 
the Floquet blocks, 
the double summation in eq.(\ref{DevLam})
can be reduced 
to a unique one in the first Brillouin zone without introducing any approximation: 
%
%
\begin{equation}
|\Psi (t)\rangle=\sum_{j=1}^{N_m}e^{-iE_{{j,0}}t/\hbar}|\lambda_{j,0}(t) \rangle \langle \lambda_{j,0}(t=0)|\Psi(t=0)\rangle
\label{Dev2}
\end{equation}
In the case of a cyclic dynamics within a $m$-dimensional subspace, 
\ref{appA} demonstrates that the sum in eq.(\ref{Dev2}) is limited to only $m$ terms. 
The wavefunction starting at $\ket{\Psi_i(0)}=\ket{i}$ can be written as
%
%
\begin{eqnarray}
\left \lbrace \begin{array}{l}
\vert \Psi_i(t) \rangle =\sum_{j=1}^m e^{-\frac{i}{\hbar}E_{j,0}t} |\lambda_{j,0}(t)\rangle U_{ji} \\
U_{ji}=\langle \lambda_{j,0}(0)|i\rangle
\end{array} \right .
\end{eqnarray}
and
the $m$ relevant Floquet eigenstates have, at the two boundaries $t=0$ and $t=T$,
non-vanishing components within the $S_o$ space exclusively, 
%
%
\begin{equation}
Q_o |\lambda_{j,0}(0)\rangle=Q_o |\lambda_{j,0}(T)\rangle=0,\; j=1,\ldots , m.
\label{Bound}
\end{equation}  
In \ref{appA} we also show that the Floquet eigenvalues $E_{j,0}$ and the $m$ components 
of the corresponding eigenstates within $S_o$ at $t=0$ can be easily derived from the time-dependent waveoperator. 

In numerical examples, the comparison of the solutions of eq.(\ref{BAEQ}) on the interval $[0,T_o]$, obtained with and without
temporal absorbing potential, 
will give us an indication about the 
naturally 
cyclic character of the dynamics driven by $H(t)$. 
The absorbing potential introduced in the interval
$[T_o,T]$  is useless if the dynamics of the wave function is purely cyclic within the
finite active subspace $S_o$. 
In what follows we will select situations and laser parameters such that controlled population exchanges
are obtained which are close to such naturally cyclic dynamics.

\section{Two illustrative examples \label{num_examples}}

\subsection{STIRAP in an asymmetric double-well}

\begin{figure}[htp]
\centering
\includegraphics[width=0.8\linewidth]{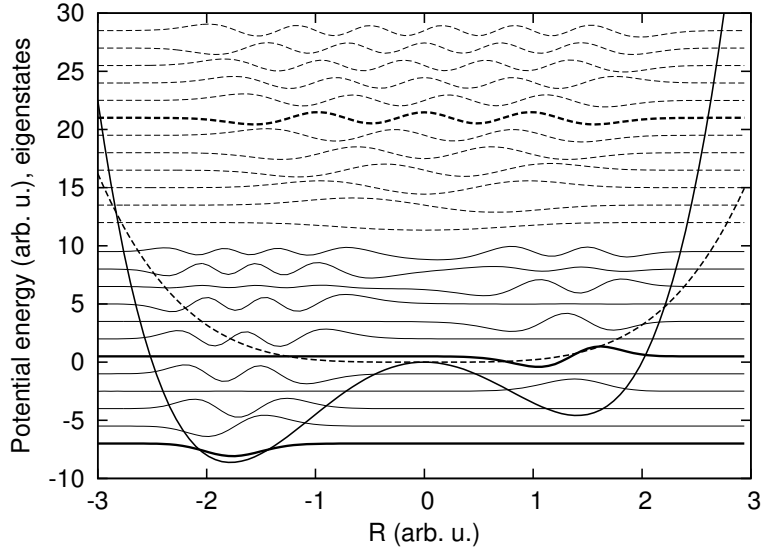}
\caption{Potential energy curves and first vibrational eigenstates
of $\epsilon_1(R)$ (full lines) and of $\epsilon_2(R)$ (dashed lines). 
The thick lines correspond to the initial, target and intermediate states. 
}
\label{Potential1}
\end{figure}

Figure \ref{Potential1} shows two potential curves for a model diatomic molecule submitted 
to two laser pulses. 
These two curves are defined as quartic polynomials, 
$\epsilon_1(R)= -5 R^2 + 0.5 R^3 + R^4 $ and
$\epsilon_2(R)= 0.2 R^4$. 
They can refer to two electronic states of a 1D vibrational Hamiltonian 
in the framework of the Born-Oppenheimer approximation. 
Similar potentials are also obtained in effective isomerization problems along a reaction coordinate \cite{Chen12}, 
or to describe the nitrogen inversion within an one dimensional method, 
the asymmetry of the potential appearing when the molecule is put down on a surface. 
In all this subsection, arbitrary units (arb. u.) are used with $\hbar=1$ 
and the various numerical parameters have been adjusted to produce realistic dynamics. 
The dipole moment which couples $\epsilon_1$ with $\epsilon_2$ is given an arbitrary constant unit value. 
%
%
%
%

The selected laser 
pulses are chosen to produce a complete transfer between the two asymmetric wells of the first surface.
In figure \ref{Potential1}, the vibrational state $(v=0,S=1)$ localized in the first well and
the state $(v=5,S=1)$ localized in the second well have in common a strong overlap with 
state $(v=6,S=2)$. 
To obtain a stimulated Raman adiabatic passage \cite{vitanov2001,Shore2008} 
the laser field 
which couples the two surfaces is chosen as the sum of two pulses 
with gaussian envelopes and carrier frequencies in resonance with the transitions $(v=0,S=1) \rightarrow (v=6,S=2)$
and $(v=6,S=2) \rightarrow (v=5,S=1)$:
%
%
\begin{equation}
E(t)=\sum_{j=1}^2 E_j \cos \left(\omega_j(t-T_j)\right) \exp\left(-\left(\frac{t-T_j}{\tau_j}\right)^2\right)
\label{Echamp}
\end{equation}
with the following numerical parameters 
%
%
\begin{eqnarray}
\left \lbrace \begin{array}{l}
E_1=0.03,\; \omega_1=4.77725153, \; T_1=250, \; \tau_1=125 \\
E_2=0.03, \; \omega_2=9.9844894, \; T_2=360, \; \tau_2=125
\end{array} \right .
\label{IterPara}
\end{eqnarray}

The nuclear dynamics is governed by the time-dependent Schr\"odinger equation within the framework
of the dipole approximation.
%
%
\begin{equation}
i \frac{\partial}{\partial t} U(t,0)=\left[ T_N +\left(
\begin{array}{cc}
\epsilon_1 & -\vec{\mu}_{1,2}.\vec{E}(t) \\
 -\vec{\mu}_{1,2}.\vec{E}(t)& \epsilon_2 
\end{array}\right) \right] U(t,0)
\label{dynbas}
\end{equation} 
where $T_N$ is the relative kinetic energy of the two atoms.

This dynamics is described within a basis set made of 
the 30 first vibrational eigenstates for each surface (namely $N_m=60$) and
$N_t=65536$ sampling time values (c.f. eq. \ref{TGRID}) 
equally distributed over the time interval $[0,T=800]$ and with a time-dependent absorbing potential
(eq. \ref{VOPT}, subsection \ref{itcalc}) localised on the time interval $[600,800]$. 
Thirty states per surface are sufficient to give convergence of the calculations with the selected
laser amplitudes and laser frequencies. 
A first calculation is made with the wave operator formalism proposed in \cite{Lecl2015}
by using a one-dimensional active space based 
on the initial state $(v=0,S=1)$. 
This choice produces a strong divergence of the algorithm from the start of 
the iterative procedure.

%
%
\begin{figure}[htp]
\centering
\includegraphics[width=0.8\linewidth]{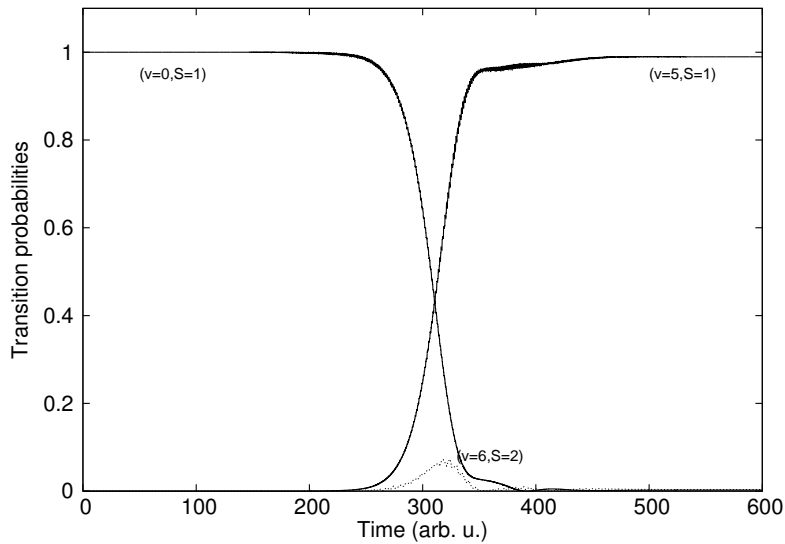}
\caption{Evolution of the populations of the initial state, $(v=0,S=1)$, the 
target state $(v=5,S=1)$ and the
intermediate state $(v=6,S=2)$ during the molecule-laser interaction.}
\label{Transfer}
\end{figure}

We then applied the formalism of the present paper by using multidimensional active spaces
of increasing dimension. 
A second attempt used an active space made up of two states:
the initial state $(v=0,S=1)$ and the target state 
$(v=5,S=1)$. 
A third attempt involves adding the intermediate state $(v=6,S=2)$ to these two states. 
Finally a fourth calculation includes the  quasi-resonant   states $(v=16,S=2)$ and $(v=6,S=1)$    
together with the three previous states
to constitute an active space $S_o$ of dimension $m=5$. 

 The second choice $(m=2)$
produces, like the first one,
a strong and rapid divergence of
the iterative calculation.
On the contrary the use of the active spaces of dimension $=3$ and $m=5$ lead to converged results, with a much better
precision in the last case $m=5$.
 The table (\ref{Table1a}) shows the convergence factor (cf eq. (\ref{IterStop}))

\begin{table}
\begin{tabular}[t]{|c|c|c|c|c|c|c|c|}
\hline Convergence
& \multicolumn{7}{c|}{Iteration number} \\
\cline {2-8}
Factor&n=1&n=2&n=3&n=4&n=5&n=6&n=7\\
\hline
$F^{(m=3)}_n$&2.11 E-02&1.58 E-03&2.09 E-04&4.89 E-04&2.86 E-04&7.78 E-04&4.21 E-04 \\
$F^{(m=5)}_n$&2.22 E-02&1.33 E-03&3.03 E-05&1.29 E-06&1.98 E-07&1.20 E-07&2.00 E-07 \\
\hline
\end{tabular}
\caption{The convergence factor $F^{(n)}_m=||\delta X^{(n)}||^2/||X^{(n)}||^2$ (eq.(\ref{IterStop}))
with respect to the iteration number $n$ for active subspaces of dimension $m=3$ and $m=5$.} 
\label{Table1a}
\end{table}

In spite of a much better convergence in the case $m=5$ than in the case $m=3$ the two calculations give
undistinguishable results in figure \ref{Transfer} with an almost complete transfer of population.
We note that the transfer is not perfectly complete ($P(t=T)_{v=0\rightarrow v=5}=0.9896$)
 and that the occupation of the 
intermediate state ($v=6,S=2$) is not strictly equal to zero during the interaction.

%
%

%
%
\begin{figure}[htp]
\centering
\includegraphics[width=0.8\linewidth]{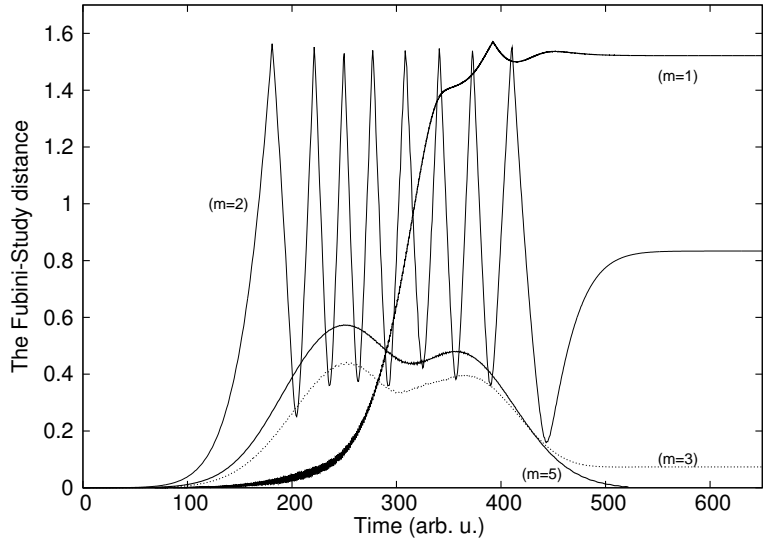}
\caption{Fubini-Study distance (\ref{FSD2}) as a function of time by using a one-dimensional active space
including only the initial state ($v=0,S=1$) (broad full line with a plateau at about $\pi/2$
for $t \geq 400$), 
a multidimensional active space of dimension $m=2$ including the
initial state and the target
state ($v=5,S=1$) (full line), 
a multidimensional active space of dimension $(m=3)$ obtained by 
adding the intermediate state ($v=6,S=2$) to the two previous states (dashed line)
and finally an active space of dimension $m=5$ by adding to the three preceding states 
the states ($v=16,S=2$) and ($v=6,S=1$) (broad full line).}
\label{Wonorm}
\end{figure}
The drastically different behaviours observed during the iterative process when the dimension $m$ changes can be understood 
by analysing the corresponding Fubini-Study distances \cite{Vien3} between
 the active spaces at the initial instant $t=0$ and
at the current time $t$ (see figure \ref{Wonorm}). 
The maximum value of $\pi/2$ associated with the notion of quantum incompability of active spaces 
is reached in the non-degenerate case ($m=1$) when the population of the initial state tends to zero, i.e. when the
complete population transfer between ($v=0,S=1$) and ($v=5,S=1$) is achieved. This limit produces the divergence of the
one dimensional wave operator, since $||\Omega (t)||=||1/\langle i|\Psi(t)\rangle ||$ and indirectly that of the
effective Hamiltonian $H_{eff}(t)=P_oH(t)\Omega(t)$ which drives the dynamics within 
the active subspace.   
On the contrary one can observe in figure \ref{Wonorm} that, in the $m=3$ and $m=5$ cases, 
the FS distance is far from its limit
 $\pi/2$ value at every time. In the last case ($m=5$), this distance even tends to very small values when
the laser is turned off. 
These small FS distances illustrate the quantum compatibility of the successive active spaces at any time and induce a fast convergence of the iterative algorithm.

%
%
\begin{figure}[htp]
\centering
\includegraphics[width=0.8\linewidth]{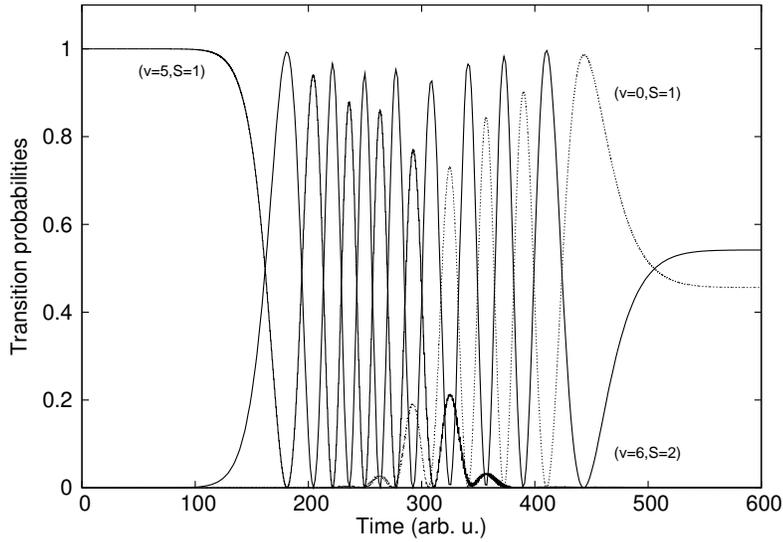}
\caption{Populations of
the three states ($v=0,S=1$), ($v=5,S=1$) and ($v=6,S=2$) when the laser field
used in the STIRAP experiment (figure \ref{Transfer}) drives a dynamics 
starting from the target state ($v=5,S=1$) (full line).
 The evolutions of  states ($v=0,S=1$) and ($v=6,S=2$) are represented by a dashed line and a broad full line, respectively.} 
\label{Evoinv}
\end{figure}
The failure in the case $m=2$ can be understood by comparing figures \ref{Wonorm} and \ref{Evoinv}. 
Because all the evolutions starting from the $S_o$ subspace are calculated as a whole, 
we shall also look at the one issuing from the target state (cf. figure \ref{Evoinv}). 
The comparison with figure \ref{Wonorm} reveals that the eight discrete time values for which the FS distance is close to $\pi/2$ 
(figure \ref{Wonorm}) 
are identical 
to the time values for which the occupations of the states ($v=0,S=1$) and ($v=5,S=1$)
decrease to zero simultaneously (figure \ref{Evoinv}). This situation is produced by the Rabi oscillations which 
affect the initial state $(v=5,S=1)$ for $t<300$ and the state $(v=0,S=1)$ for $t>300$ in this case. 
As a consequence, the $(2 \times 2)$ matrix $P_o^{\dagger} [ U(t,0;H)] P_o$ which leads to 
the FS distance (equation \ref{FSD2}),
exhibits, at these eight discrete time values, a column
corresponding to the initial state ($v=5,S=1$) equal to zero and consequently a FS distance close to $\pi/2$. 
Figure \ref{Evoinv} shows that other Rabi oscillations affect the intermediate state ($v=6,S=2$) but the zero values of these
oscillations correspond to maximum values of the oscillations affecting the initial state ($v=5,S=1$) and the state ($v=0,S=1$). This
explains why correct results with small FS distances are obtained by using an active space of dimension $m=3$.

The best choice for the active space is the one with dimension $m=5$.  
This choice induces FS distances wich remain very small at all times. 
Moreover these distances converge to very small values when the laser field is turned off (figure \ref{Wonorm}),
 indicating that the wave function is, at the end, projected onto the initial $S_o$ subspace. 
In other words the dynamics is approximately cyclic within this 5-dimensional subspace. 
The selection rules used to build the subspace are simple. The subspace should include the initial and
the target state, namely ($v=0,S=1$) and ($v=5,S=1$) in the STIRAP experiment. 
The subspace should also include states which are strongly coupled to these first two ones by near resonant effects,
$E(v=6,S=2)-E(v=0,S=1) \simeq \hbar \omega_2$ and $E(v=16,S=2)-E(v=5,S=1) \simeq \hbar \omega_2$. 
Finally the state $(v=6,S=1)$, which is not in exact resonance,
has been added to these first four states.  
It is weakly populated during the interaction. 
In the one-dimensional subspace case, 
the divergence of the wave operator is related to the complete population transfer between
$(v=0,S=1)$ and $(v=5,S=1)$. 
This also produces a divergence in the effective Hamiltonian $H_{eff}(t)=P_o H(t) \Omega (t)$ which
drives the dynamics within the active subspace. 
On the contrary, the use of an active subspace of dimension $m=5$, induces 
a $(5 \times 5)$ $H_{eff}$ matrix whose components are always finite. 
%
%
%
%
\begin{figure}[htp]
\centering
\includegraphics[width=0.45\linewidth]{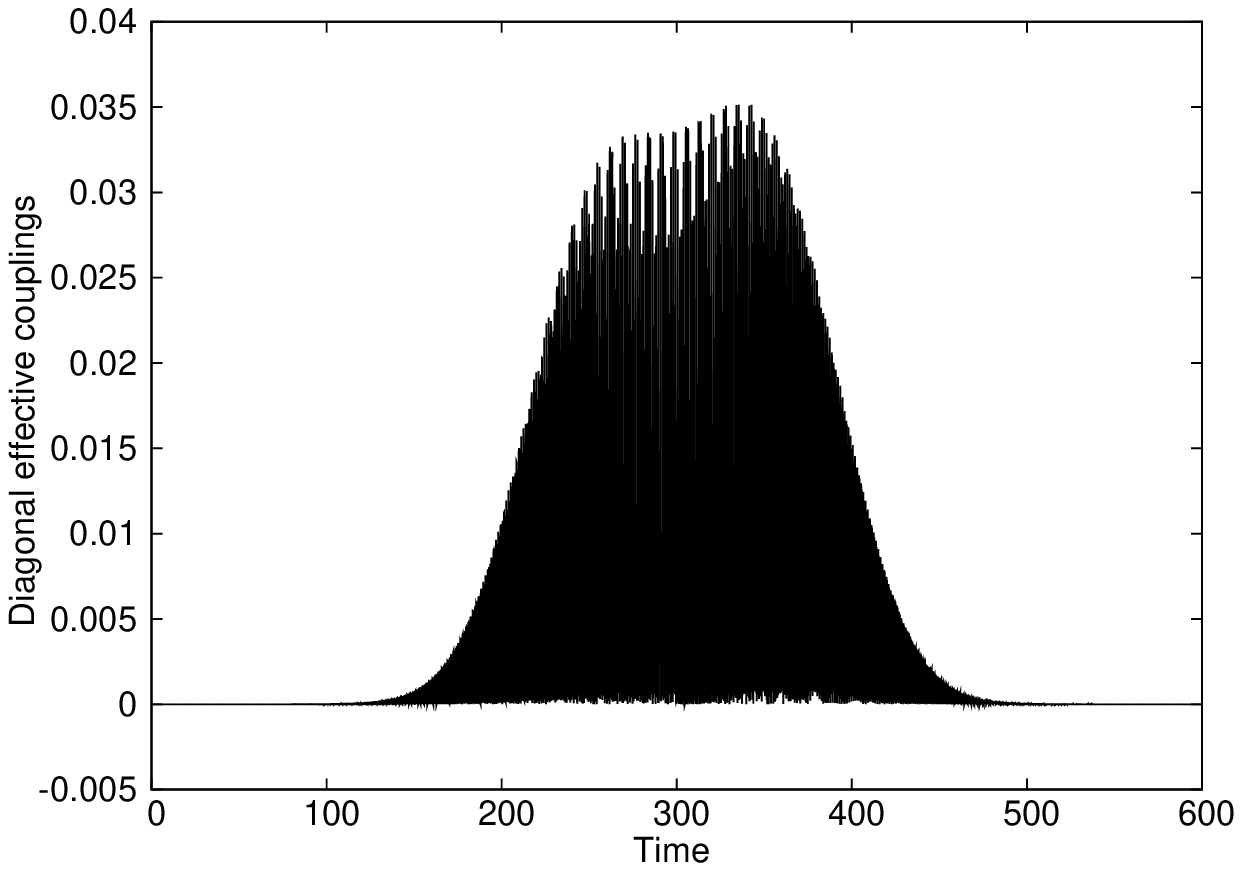}
\includegraphics[width=0.45\linewidth]{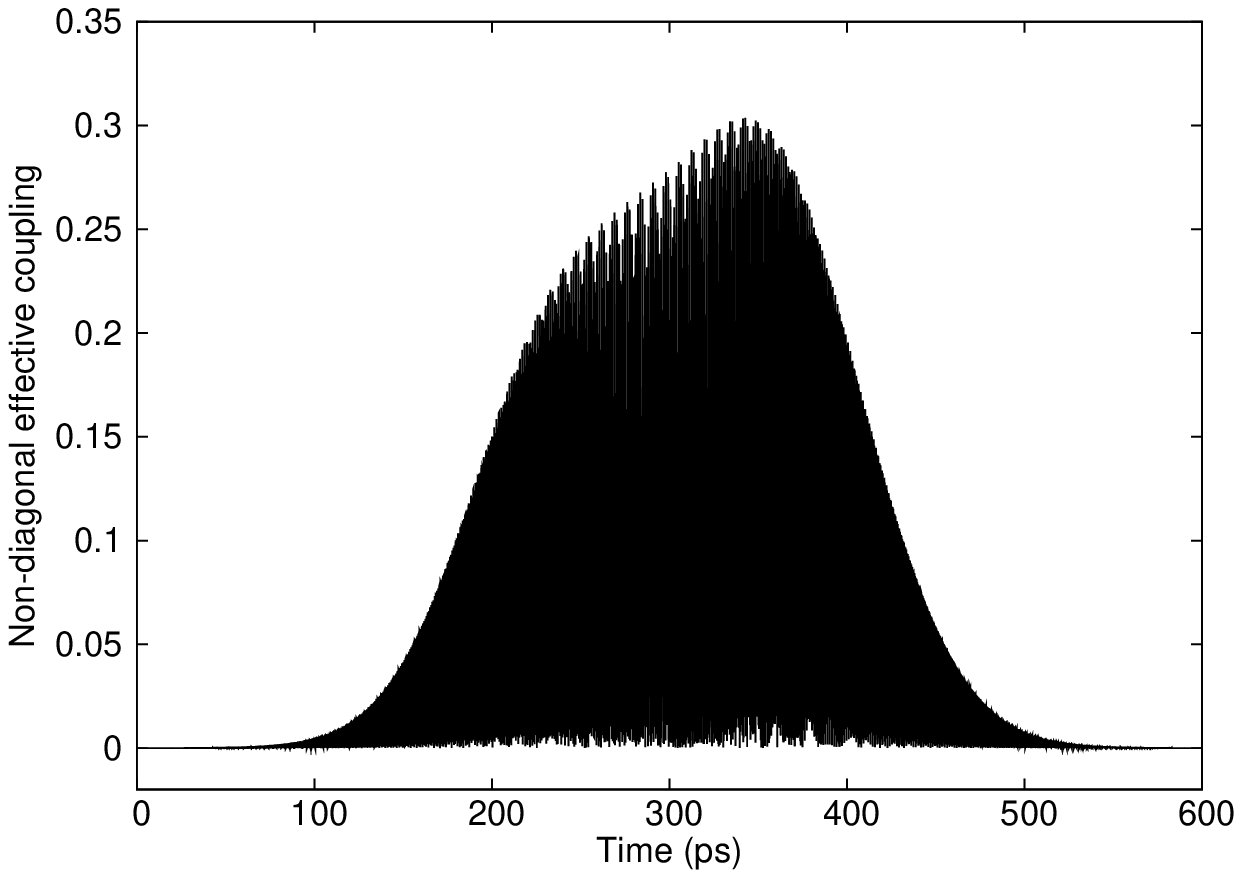}
\caption{Shift from the non-perturbed values of the 
diagonal matrix elements (modulus) of $H_{eff}$, for the initial state $(v=0,S=1)$ (left frame). 
The right frame represents the coupling amplitude between states $(v=5,S=1)$ and $(v=6,S=2)$ }
\label{HEFFd}
\end{figure}
The diagonal element showing the largest shift is the one   
which corresponds to the initial state $(v=0,S=1)$ and it is presented in 
figure (\ref{HEFFd}). It exhibits relatively small modulations
compared with the spacing between the non-perturbed eigenvalues, 
$E(v=5,S=1)-E(v=0,S=1) \simeq 5.207 $ arb.u. and $E(v=6,S=2)-E(v=0,S=2) \simeq 9.984 $ arb.u. 
Nevertheless correct solutions are obtained if, and only if
these modulations are taken into account during the calculation. 
 The non-diagonal couplings are much larger, 
especially the direct couplings $(v=0,S=1) \leftrightarrow (v=6,S=2)$ and $(v=6,S=2) \leftrightarrow (v=5,S=1)$ (see figure (\ref{HEFFd})).
The effective
Hamiltonian $H^{eff}$ also possesses a small direct coupling between the initial $(v=0,S=1)$ and the final state $(v=5,S=1)$
(not shown).
%
%

%
%
\begin{figure}[htp]
\centering
\includegraphics[width=0.45\linewidth]{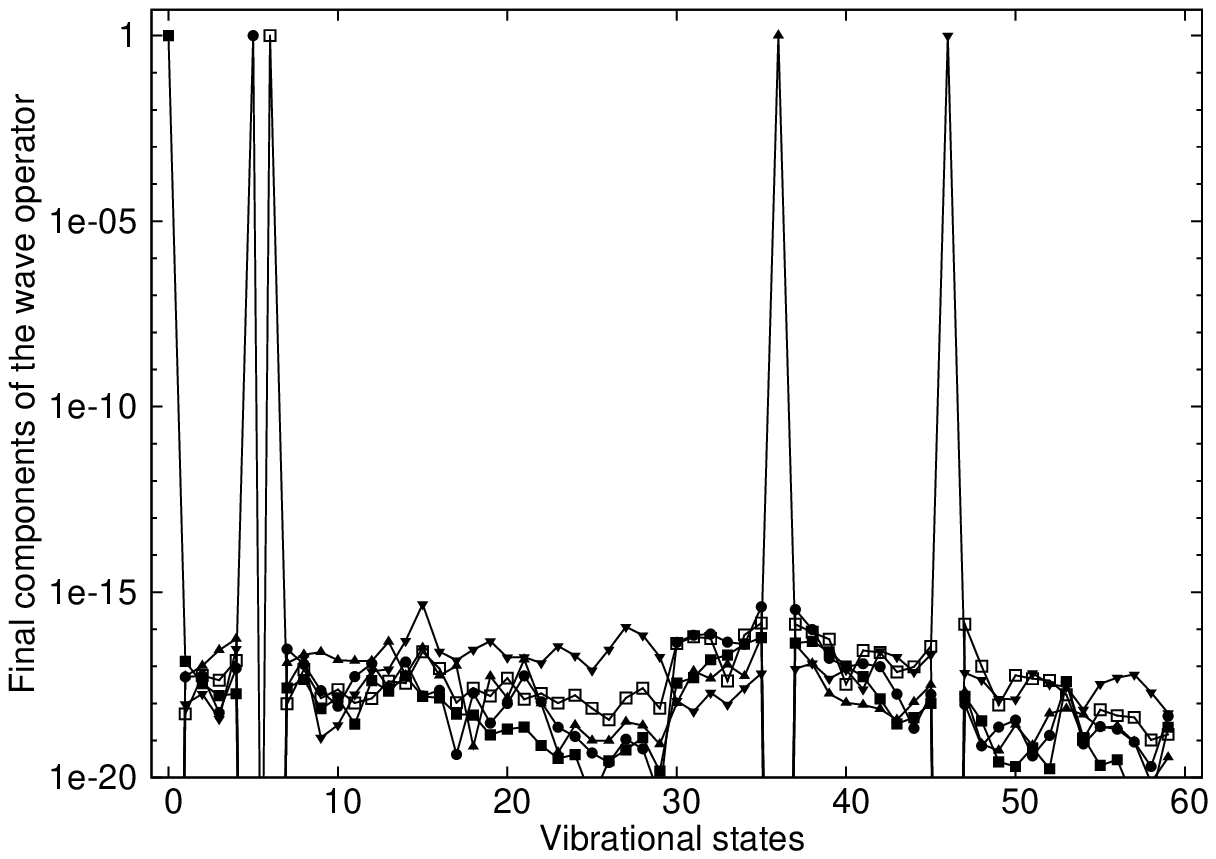}
\includegraphics[width=0.45\linewidth]{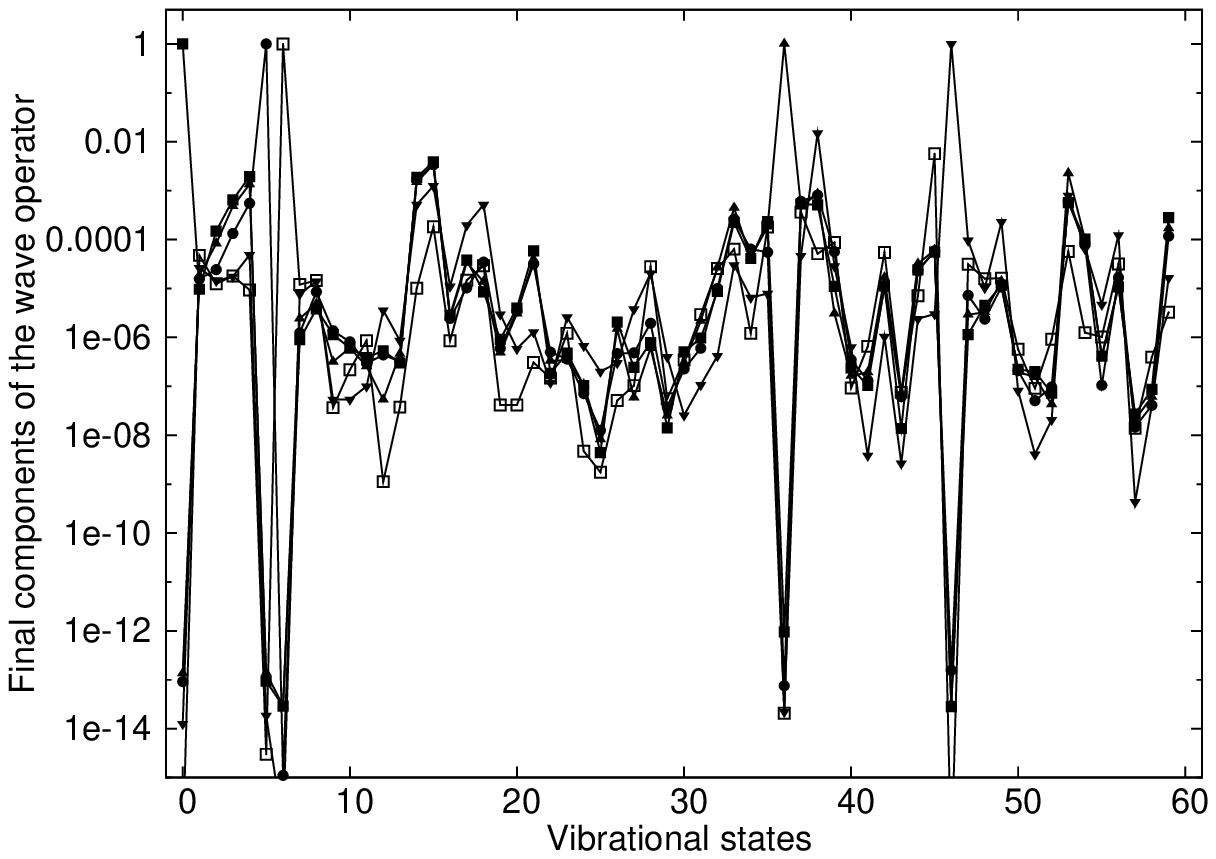}
\caption{Component amplitudes of the five columns which constitute the wave operator at the final time $(t=T)$
when equation (\ref{BAEQ}) is integrated with time-dependent absorbing potential (left frame) and without
time-dependent absorbing potential (right frame). 
Five different symbols are
associated to the five 
initial
states which compose the active space $S_o$, 
$\blacksquare \leftrightarrow (v=0,S=1)$,
$\bullet \leftrightarrow (v=5,S=1)$, 
$\blacktriangle \leftrightarrow (v=6,S=2)$
$\blacktriangledown \leftrightarrow (v=16,S=2)$, 
$\square \leftrightarrow (v=6,S=1)$}
\label{Cycl1}
\end{figure}

We can analyse the cyclic character of the dynamics with respect to the selected  active space with $m=5$. 
For doing this analysis we need to remove the artificial influence of the time-dependent absorbing potential defined in equation \ref{VOPT}. 
For a purely cyclic dynamics the solutions of eq. (\ref{BAEQ}) obtained with and without time-dependent
absorbing potentials would be strictly equal.  
The wave operator components (obtained with and without artificial absorbing potential) are shown in figure \ref{Cycl1}.
Vibrational states are numbered from $v=0$ to $v=29$ for the first surface and from $v=30$ to $v=59$ for the second surface. 
In both frames of figure \ref{Cycl1}, 
the five unit peaks correspond to the return of the wave operator to the initial active space at the end of the dynamics.  
The projection of $\Omega$ into the active space at $t=T$ is effectively identical to $P_o$ (the zero terms being approximated by small values of about $10^{-13}$). 
However 
in the right frame 
the ($5 \times 55$) non-diagonal components 
coupling the active space to the complementary space
take non-vanishing small values between $10^{-2}$ and $10^{-6}$, which corresponds to 
transition probabilities smaller than $10^{-4}$. 
This figure proves that the dynamics is largely (but not perfectly) cyclic
within the selected 5-dimensional active space. 
In such a case the solution can be expanded on a basis set made from only five periodic Floquet eigenstates associated with the active space. The five eigenvalues and the initial components of the corresponding eigenvectors have been calculated using the method explained in \ref{appA} and are given in table \ref{Table1}. These Floquet eigenstates have non-vanishing components at the two time-boundaries in the active space exclusively. 
Note that these vectors and the corresponding eigenvalues depend on the Floquet Hamiltonian but also on the duration $T$ of the selected time interval.

\begin{table}
\begin{tabular}[t]{|c|c|c|c|c|c|}
\hline
& \multicolumn{5}{c|}{Floquet eigenvalues $E_{\lambda_j}$} \\
\cline {2-6}
&5.6692 E-04&-2.8759 E-03&-2.7546 E-03&-1.42049 E-03&-1.6374 E-03 \\
\hline
\hline
$\langle v,S|$ & \multicolumn{5}{c|}{Eigenvector components 
$|\langle v,S|\lambda_j (t=0) \rangle|$ }\\
\hline
(v=0,S=1)&0.6531&0.6129&0.2282&0.3110&20376 E-02 \\
(v=5,S=1)&0.6616&0.6106&0.1970&0.3875&2.0543 E-02 \\
(v=6,S=2)&0.3674&0.3765&0.1445&0.8363&5.9142 E-02 \\
(v=16,S=2)&2.4062 E-02&0.3321&0.9424&2.9068 E-02&1.9696 E-03 \\
(v=6,S=1)&7.2863 E-03&4.0218 E-03&1.3086 E-03&6.5247 E-02&0.9978 \\ 
\hline
\end{tabular}
\caption{The five Floquet eigenstates $\vert \lambda_j \rangle$ 
over which the cyclic wavefunctions can be expanded and the corresponding eigenvalues. } 
\label{Table1}
\end{table}

The small defect with respect to a perfect cyclicity obtained in the waveoperator components 
is consistent with the results for the FS distance previously shown in figure \ref{Wonorm} 
for an active space of dimension $m=5$. 
At time $t=600$, the laser is turned out and the FS distance is about $10^{-2}$ and not strictly equal to zero, 
indicating that a small part of the population is present in the complementary space.

\subsection{Dissociative dynamics of $H_2^+$}

\begin{figure}[htp]
\centering
\includegraphics[width=0.8\linewidth]{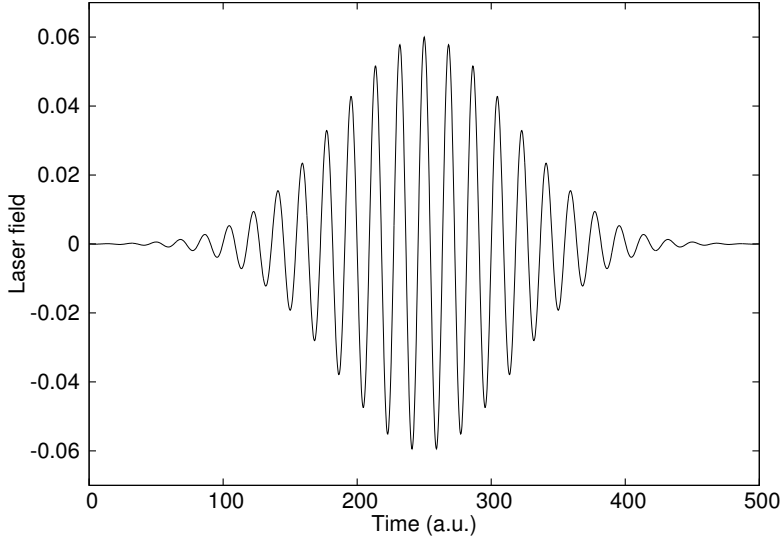}
\caption{Laser field amplitude as a function of time} 
\label{LASERFIELD}
\end{figure}

The second illustrative example is that of the $H_2^+$ molecule submitted to an intense laser pulse.
The principal aim of this example is to test the ability of the global algorithm 
to describe non-adiabatic dynamics driven by a non-hermitian Hamiltonian. 
We only take into account the two first effective potentials \cite{Bunkin} corresponding to 
the two lowest electronic states $^2\Sigma_g^+$ and $^2\Sigma_u^+$.
We make the assumption that the rotational dynamics is frozen. 
This is a sensible assumption because we consider only very short laser pulses. 
Before calculating the dynamics, the field-free Hamiltonian of $H_2^+$ has been pre-diagonalized  
on a radial grid basis using a grid method  with a radial complex absorbing potential \cite{Marston1989, Poirier}. 
A non-perturbed vibrational eigenbasis made of 
$2 \times 200 $ eigenvalues $\varepsilon_j$ and bi-orthogonal eigenstates $\{|j\rangle, |j^*\rangle\}$    
is then used 
(see \ref{appnonherm}) 
to express the electric dipole moment operator  and the corresponding
matrix $\mu_{ij}$ which couples the two surfaces. 
Within this simple Born-Oppenheimer model, the lower surface supports  $N_{BS}= 19$ bound vibrational states. 
The electric field is a sum of two slightly detuned simultaneous pulses
(the detuning has been adjusted to correspond to the spacing between the first two vibrational states, 
$\omega_1 - \omega_2 \simeq E(v=1)-E(v=0)$). 
We use again the expression (\ref{Echamp}) 
with the following parameters, given in atomic units:
%
%
\begin{eqnarray}
\left \lbrace \begin{array}{l}
E_1=E_2=0.03 \; a.u.,\; \omega_1=0.35 \; a.u., \; \omega_2=0.3398 \; a.u.,  \\
T_1=T_2=250 \; a.u.,  \;\tau_1=\tau_2=100 \; a.u.
\end{array} \right .
\label{IterPara2}
\end{eqnarray}
This laser pulse is represented in figure (\ref{LASERFIELD}). 
The peak value equal to $E=0.06$ a.u. corresponds to an
intensity of $I=1.263 \times 10^{14}$ W/cm$^2$.
An ample Fourier basis set of $N_t=2048$ functions is used to represent the interaction throughout the time interval $[0,640 \;au]$, which is widely sufficient to include all the possible multiphoton processes expected with fields exceeding $10^{14}$ W/cm$^2$ . 

%
%
The dynamics is integrated by using an initial active space of dimension $m=41$ which includes 
the $N_{BS}=19$ bound states of the first surface $^2\Sigma_g^+$, 
the 11 first pseudo-diffusion states which discretize the continum of this surface 
and the 11 last pseudo-diffusion states which discretize the continuum of the second surface $^2\Sigma_u^+$. 
The dimension of this subspace is important but still small 
compared with the dimension of the molecular basis (here $N_m = 400$).
Why do we precisely choose this active subspace~? 
Including all the bound states in the active space is certainly a good choice. The initial wavepacket will be in general chosen as a bound wavepacket
 and after the pulse is turned off, the remaining bound wavepacket returns to this subspace. 
The active subspace can also be completed by including some of the discretized continuum states. 
The selection has been done by looking at the relative lifetimes of the different pseudo-diffusion states.  
States with long lifetimes may contribute to the final wavepacket and have been included in the active subspace.
There are diffusion states with long lifetimes close to the dissociation limit of the first potential curve $^2\Sigma_g^+$ 
and other long lifetime states close to the energy trucation of the second curve $^2\Sigma_u^+$. 
The complementary space is thus made of all the other pseudo-diffusion states with shorter lifetime, leading to 
molecular photodissociation.


%
\begin{figure}[htp]
\centering
\includegraphics[width=0.8\linewidth]{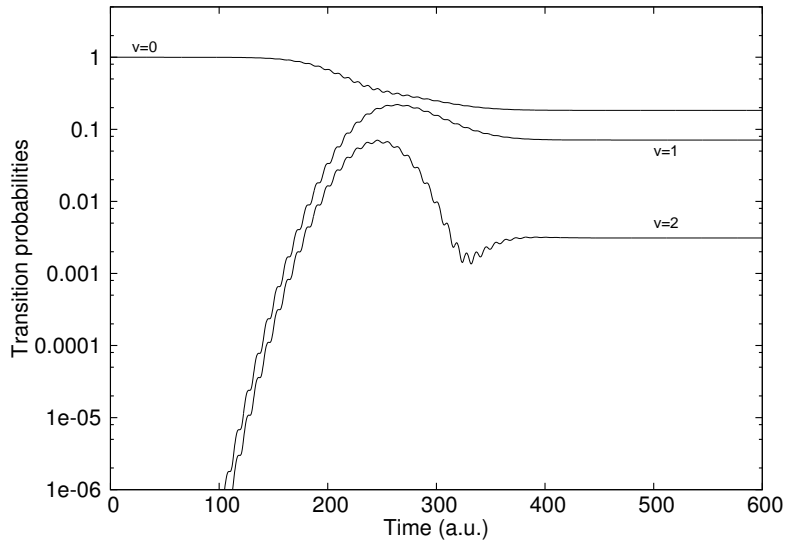}
\caption{Transition probabilities from the initial state $(v=0)$
as a function of time, 
$P(v=0\rightarrow v=0)$, 
$P(v=0\rightarrow v=1)$, 
$P(v=0\rightarrow v=2)$.}
\label{PROBAH20}
\end{figure}
%
%

%

%
\begin{figure}[htp]
\centering
\includegraphics[width=0.8\linewidth]{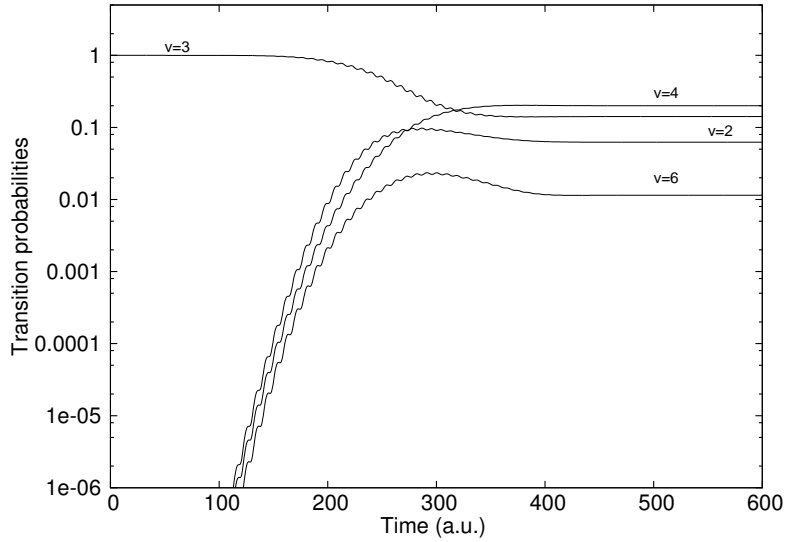}
\caption{Transition probabilities from the initial state $(v=3)$
as a function of time,
$P(v=3\rightarrow v=2)$, 
$P(v=3\rightarrow v=3)$,  
$P(v=3\rightarrow v=4)$, 
$P(v=3\rightarrow v=6)$.} 
\label{PROBAH23}
\end{figure}
By including all the bound states in $S_o$, we integrate globally the transition dynamics between these bound states so that the wave operator obtained
 by solving eq. (\ref{BAEQ}) gives us all the transition probabilities: 
$P_{i\rightarrow j}(t); \;
i=0 \ldots 18, j=0 \ldots 18$. 
Two partial results are given in figures (\ref{PROBAH20}) and (\ref{PROBAH23}) for
dynamics issuing from states $(v=0)$ and $(v=3)$, respectively. 
Moreover the various dissociation probabilities $P_{diss}(i)$, 
for an evolution issuing from the initial state $v=i$ can be obtained as 
%
\begin{equation}
P_{diss}(i)=1-\sum_{j=0}^{18} P_{i \rightarrow j}(t=T)
\label{H2DISS}
\end{equation}  
The iterative procedure (eqs \ref{Sol1}- \ref{Incr})  converges after only $n=14$ iterations
 to $\epsilon=3. 10^{-8}$, 
 giving all the 
$(N_{BS} \times N_{BS})$ 
transition probabilities with an accuracy 
of four stable digits.

\begin{table}
\begin{tabular}[t]{|c|c|c|c|c|c|c|c|}
\hline 
& \multicolumn{7}{c|}{Convergence factor} \\
\hline
Iteration number&n=1&n=2&n=3&n=4&n=5&n=6&n=7\\
\hline
$F^{(m=41)}_n$&5.08 E-03&6.01 E-04&1.02 E-03&6.81 E-04&2.47 E-04&2.75 E-04&1.27 E-04\\                   
\hline
Iteration number&n=8&n=9&n=10&n=11&n=12&n=13&n=14\\
\hline
$F^{(m=41)}_n$&5.43 E-05&3.87 E-05&1.32 E-05&3.49 E-06&7.32 E-07&1.54 E-07&3.03 E-08 \\
\hline
\end{tabular}
\caption{The convergence factor $F^{(n)}_m=||\delta X^{(n)}||^2/||X^{(n)}||^2$ (eq.(\ref{IterStop}))
with respect to the iteration number $n$ for the active subspace of dimension $m=41$.} 
\label{Table2a}
\end{table}

The table (\ref{Table2a}) shows the convergence factor (cf eq. (\ref{IterStop})).
This good result is understandable since the active space includes all the  
bound states which mainly participate in the dynamics.
It is true that a large part of the wave packet is projected into the two continua. 
But most of the pseudo-diffusion states which span these two continua have small lifetimes and their populations
rapidly decrease to zero. 
Some long-lived diffusion states are present and can disturb the cyclic character of the dynamics but this defect is
suppressed in the present treatment by including these states in our active subspace.
As expected the populations of the long-lived diffusion states included in the active subspace states do not converge to zero at $t=T$. 
These results are confirmed by testing the convergence versus the composition of the active space. 
Including all the bound states is essential in this rather non-adiabatic example. This is the safest way to ensure an easy convergence for any dynamics issuing from bound states (for example, selecting a too small active space of dimension $m=5$ made of the first five bound states $v=0$ to $v=4$ is not sufficient and makes the iterative procedure diverge). 
In table \ref{Table2a} which corresponds to $m=41$, the convergence 
factor, equal to $5.08 \times 10^{-3}$ for $n=1$, converges to $3.03 \times 10^{-8}$ after $n=14$ iterations. 
By reducing the active space to the $m=19$ bound states,
the convergence factor converges from $2.84 \times 10^{-3}$ for $n=1$ to $3.33 \times 10^{-7}$ for $n=14$. 
This worse result is due to non-negligible populations of some long-lived diffusion states which subsist at the end of the interaction and are worth being included in the active subspace. 
Increasing $n$ from $41$ to $51$ by adding more diffusion states does not significantly affect the results of table \ref{Table2a}. 
 
%
%
%
%
\begin{figure}[htp]
\centering
\includegraphics[width=0.8\linewidth]{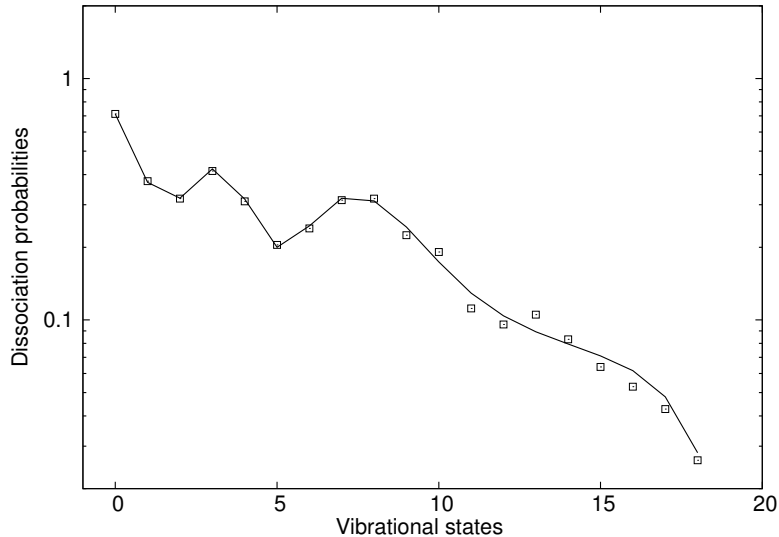}
\caption{Dissociation probabilities associated to the initial states $(v=0, \ldots v=18)$,
obtained by integrating the equation (\ref{BAEQ}) in the presence of a time-dependent absorbing potential
(continuous line), 
and without absorbing potential ($\square$). } 
\label{PDISS}
\end{figure}

\begin{figure}[htp]
\centering
\includegraphics[width=0.8\linewidth]{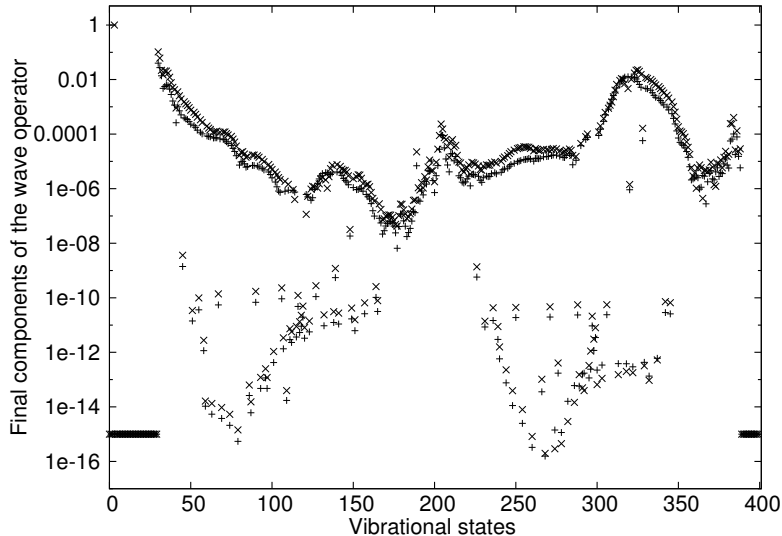}
\caption{Wave operator component amplitudes, for the columns corresponding to the initial states $v=0$ ($+$) and $v=3$ ($\times $) 
at the final time $(t=T)$ when the equation (\ref{BAEQ}) is solved without absorbing potential. 
The laser field is built on the interval $[0,500]$ a.u. and the final time is $T=640.$ a.u.} 
\label{FIN512}
\end{figure}

Figure (\ref{PDISS}) displays the dissociation probabilities expressed in eq. (\ref{H2DISS}). 
The nearly equal results obtained with and without absorbing potentials confirm that the dynamics is approximately cyclic
with respect to the selected active subspace. 
This character is also confirmed by analysing the figure (\ref{FIN512})
which represents the amplitudes of the components of $X=Q_o \Omega$ at the final time $(t=T)$.
All these components are smaller than $10^{-1}$, most of them being between $10^{-2}$ and $10^{-6}$. 
%
%

\section{Conclusion \label{conclusion}}
The two illustrative examples presented in section \ref{num_examples} leads to the following conclusions.
The multidimensional version of the global integrator significantly improves the performances 
of the previous one-dimensional integrator of ref. \cite{Lecl2015}. 
If the active space is correctly chosen, the divergences appearing in the one-dimensional case 
disappear and correct solutions of the Schr\"odinger equation with initial conditions
 corresponding to each one of the unperturbed molecular states which compose
the initial active space are obtained, by solving only once the equation (\ref{BAEQ}). 
In these ideal situations the convergence is fast and accurate solutions are obtained after only a few iterations.

The present theory treats both the
periodic and the quasi periodic perturbations arising in the interactions between matter and c.w. laser fields or pulsed
laser fields. It is applicable to the non-Hermitian Hamiltonians generated by using 
analytical continuations of the resolvent in the complex plane when $L^2$ representations of the continua are used. 
Moreover the global character of the integration algorithm makes possible the rapid repetition 
of perturbative calculations when some of the physical parameters (intensity, wavelength) are slightly modified. 
This feature is well adapted to investigate control processes which call for repeated propagation attempts.

The most delicate point is the selection of a good initial active space. 
Several concepts which are closely related 
(the cyclic character of the dynamics, 
the Fubini-Study distance, 
the time-dependent absorbing potential) 
participate in the selection of such a 
good active space, which should be as small as possible while giving a rapid convergence.
In practice, looking at the coupling strength between states, including resonant and near-resonant states and long-lived continuum states, is a good
guide in selecting the active space.
The quality of the active space (size and choice of the relevant states) has direct consequences on the numerical convergence
and can also be appreciated by monitoring a posteriori the Fubini-Study quantum distance 
between the initial, fixed subspace and the dynamical subspace. 
For example the iterative treatment diverges
if this FS distance tends to $\pi/2$ at a given arbitrary instant between $t=0$ and $t=T$. 

The use of Floquet theory necessitates that the dynamics is cyclic. 
We have shown that (see \ref{appA}) in such a case (i.e. if the wave function included 
in the active space $S_o$ of dimension $m$ at $t=0$ returns to this subspace at the final time $T$), 
the wave function can be rigorously expanded on a very small 
Floquet eigenbasis set of dimension $m$. This is an important
result which will be exploited in later work. 
Unfortunately, a spontaneous cyclic character is never rigorously observed. 
An artificial cyclic dynamics can be obtained by using a generalization of the 
constrained adiabatic trajectory method, 
i.e. by introducing an asymptotic time-dependent absorbing potential (eq.~\ref{VOPT}) in order to impose the condition $X(t=T)=0$. 
The absorbing potential imposes the periodicity
and suppresses the problems near the boundary $t=T$ (except in the case of a complete inversion
with a one-dimensional active space). However it does not suppress the
divergences appearing when the FS distance tends to $\pi/2$ at intermediate instants. 
The best scheme for the selection of the active space 
is to work first without an absorbing potential by using simple selection criteria
and then to add the absorbing potential at a second stage to suppress the small inconsistencies due to the non-perfect cyclicity. 
The selection criteria should take into account the distribution on the energy scale of the eigenstates dressed by the laser field.

The second illustrative example shows that our algorithm works well 
for dissipative systems when grouping together all the coupled bound states within the active space. 
In the H$_2^+$ case the global integrator dresses all the $N_{BS}=19$ bound states with the 
continua and transforms them into $N_{BS}$ resonance states which mainly participate in the dissociative dynamics.

\ack
Simulations have been executed on computers of the Utinam Institute of the Universit\'e de Franche-Comt\'e, supported
by the R\'egion de Franche-Comt\'e and Institut des Sciences de l'Univers (INSU).

\appendix
\section{Cyclic evolution and Floquet basis set \label{appA}}

In this appendix we prove that imposing the cyclicity of the wavefunction with respect to a fixed $m-$dimensional subspace $S_o$ implies that
 only $m$  Floquet eigenvectors participate in its development (\ref{Dev2}), those vectors having non-zero components only in $S_o$ at $t=0$ (hence at $t=T$).

\subsection{Cyclic wavefunction}

Let $S_o$ be a subspace of dimension $m$ of the Hilbert space with projector
%
%
\begin{equation}
P_o=\sum_{j=1}^m |j\rangle \langle j| 
\label{AProj}
\end{equation} 
and let $H(t)$ be the time-dependent hermitian Hamiltonian which drives the dynamics of the wave function, starting from the $S_o$ space, 
over the time interval $[0,T]$. 
If the evolution of the wavefunction is cyclic with respect to $S_o$ and if $\{|i\rangle\} \; i=1,\ldots m$ is
a complete basis of this subspace then:
%
%
\begin{equation}
\forall i\leq m \;\quad  \vert \Psi_i(0) \rangle =|i\rangle \Rightarrow \vert \Psi_i (T) \rangle=\sum_{j=1}^m \Phi_{ji}|j\rangle
\label{Acycl}
\end{equation}
where $\Phi$ is a unitary matrix of dimension $m$
(the basis set can be composed of the eigenvectors of the molecular Hamiltonian, $H_o \vert i \rangle = e_i \vert i \rangle$). 
If $U$ is the matrix which diagonalize the unitary matrix $\Phi$, namely
%
%
\begin{equation}
\tilde{\Phi}=U \Phi U^{-1},\; \mathrm{ with } \; \tilde{\Phi}_{jj}=e^{i\phi_j}
\label{ADiag}
\end{equation}
with $\phi_j$ real, then by introducing the new basis set $|\tilde{k}\rangle=\sum_{j=0}^m \bar{U}_{kj}|j\rangle$ 
(the bar denoting the complex conjugate), one obtains
%
%
\begin{equation}
\vert \Psi_i(T) \rangle =\sum_{k=1}^m e^{i \phi_k} U_{ki}|\tilde{k}\rangle. 
\label{AExpa}
\end{equation}
\subsection{Expansion on the Floquet basis set}
The total interval $T$ is seen as a period for periodic Floquet eigenvectors. 
We assume that the Floquet spectrum is non degenerate. 
Using eqs.(\ref{Dev2}) and (\ref{AExpa}), one can write
%
%
%
\begin{equation}
\vert \Psi_i(T) \rangle =\sum_j \sum_{k=1}^m e^{-\frac{i}{\hbar}E_{j,0}T} |\lambda_{j,0}(0)\rangle \langle \lambda_{j,0}(0)|\tilde{k}
\rangle U_{ki}
\label{AExpb}
\end{equation}
Using eq. (\ref{AExpa}) and projecting on $\langle \tilde{k} \vert$ gives
%
%
\begin{equation}
\langle \tilde{k}|\Psi_i(T)\rangle=e^{i\phi_k} U_{ki} \quad \forall k \leq m
\label{Aproj2}
\end{equation}
Introducing (\ref{AExpb}) into (\ref{Aproj2}) leads to 
%
%
\begin{equation}
\sum_j e^{-\frac{i}{\hbar}E_{j,0}T -i\phi_k}\langle \tilde{k}|\lambda_{j,0}(0)\rangle\langle \lambda_{j,0}(0)|
\tilde{i}\rangle = \delta_{ki}
\label{APrinc}
\end{equation}
By using the following notations,
%
%
\begin{eqnarray}
\left \lbrace \begin{array}{l}
d_{ji}=\langle \lambda_{j,0}(0)|\tilde{i}\rangle \\
\phi_{jk}=-\frac{1}{\hbar}E_{j,0}T-\phi_k  \; 
\end{array} \right .
\label{ADef}
\end{eqnarray}
and using the closure relation on the $\vert \lambda_{j,0} \rangle$, 
eq.(\ref{APrinc}) leads in the case $i=k$ to the constraint
%
%
\begin{equation}
\sum_j |d_{ji}|^2 (1-e^{i\phi_{ji}})=0.
\label{ARes0}
\end{equation}
In each term of this sum $|d_{ij}|^2$ is a positive real number and $1-e^{i\phi_{ji}}$ is a complex number
localized in the half plane $x >0$, 
on a circle tangent to the vertical axis and passing through zero only if $\phi_{ij}=0$. 
Consequently eq.(\ref{ARes0}) can be satisfied only if there exist some $j=j_i$ such that
%
%
\begin{eqnarray}
\left \lbrace \begin{array}{l}
 d_{j_i i}\not= 0 \; \mathrm{and}\; \phi_{j_i i}=0 \; \Rightarrow \phi_i= \frac{1}{\hbar} E_{j,0} T \\
d_{ji}=0, \;\; \forall j \not= j_i. 
\label{ARes}
\end{array} \right .
\label{ACond}
\end{eqnarray}
If two or more terms among the $d_{ji}$ were different from zero, this would imply that the corresponding phases
$\phi_{ji}$ are simultaneously zero, which is impossible since we have assumed a non-degenerate Floquet spectrum. 

In the case $i \not= k$, 
eq.(\ref{APrinc}) reads
\begin{equation}
\sum_j e^{- i \phi_{jk}} \bar{d}_{jk} d_{ji} = 0
\label{Aikdiff}
\end{equation}
Considering eq.(\ref{ACond}), we see that $d_{j k} = 0$ except when $j=j_k$, and 
the same is true for $d_{ji}=0$ except when $j=j_i$. Then eq. (\ref{Aikdiff}) implies that
%
%
\begin{equation}
i \not= k \; \Rightarrow \; j_i \not= j_k.
\label{ARes2}
\end{equation}
Finally the results (\ref{ARes}) and (\ref{ARes2}) prove that only $m$ periodic Floquet eigenvectors participate
 in the wave function expansion with the conditions $P_o |\lambda_{j_i,0}(0)\rangle=|\lambda_{j_i,0}(0)\rangle$.
By introducing a new numbering of the basis set such that $j_i \rightarrow i$ we obtain
%
%
\begin{eqnarray}
\left \lbrace \begin{array}{l}
\vert \Psi_i(t) \rangle =\sum_{j=1}^m e^{-\frac{i}{\hbar}E_{j,0} t} |\lambda_{j,0}(t)\rangle U_{ji} \\
U_{ji}=\langle \tilde{j}|i\rangle=\langle \lambda_{j,0}(0)|i\rangle
\end{array} \right .
\label{AFin}
\end{eqnarray}

\subsection{Generalized Floquet state components from the wavefunction} 

Once the evolution operator is obtained, it is possible to deduce the initial and final 
components of the $m$ generalized Floquet vectors which participates in the $\Psi_i$ expansion.
By using the T-periodicity of Floquet eigenvectors, (\ref{AFin}) at initial and final time $T$ gives
\bea
P_o \ket{\Psi_i(0)} = \ket{i} = \sum_{j=1}^m P_o \ket{\lambda_{j,0} (0) } U_{ji} \nonumber \\
P_o \ket{\Psi_i(T)} = \sum_{j=1}^m e^{-\frac{i}{\hbar} E_{j,0} T} P_o \ket{\lambda_{j,0} (0) } U_{ji}.
\eea
Using $(m\times m)$ matrices with $\mathbf{\Psi}_{ki}=\langle k \ket{\Psi_i(T)}$,
$\mathbf{\Lambda}_{ki}=\langle k \ket{\lambda_{i,0}(0)}$, 
$\mathbf{E}_{ki} = e^{-\frac{i}{\hbar}E_{k,0}T} \delta_{ki}$, $U$ defined in (\ref{ADiag})
and the identity matrix $I$, 
we obtain 
\beq
\left \lbrace \begin{array}{l}
\mathbf{\Psi} = \mathbf{\Lambda} \mathbf{E} U \\
I = \mathbf{\Lambda} U
\end{array} \right.
\label{Lambcomp}
\eeq
which leads to the following result:
\beq
\mathbf{\Psi} = \mathbf{\Lambda E \Lambda}^{-1}.
\eeq
Initial (and final) components of the $m$ generalized Floquet states of interest within the subspace $S_o$
and the associated Floquet eigenvalues can then be calculated by diagonalizing the small matrix $\mathbf{\Psi}$.

\subsection{Non-hermitian case \label{appnonherm}}
The above reasoning can be followed also in the case of a non-hermitian Hamiltonian
accounting for dissipative system such as photodissociation problems. 
Biorthonormal basis sets with left and right eigenvectors must be introduced for both the molecular Hilbert space,
\bea
H_0 \vert j \rangle = \varepsilon_j \vert j \rangle, \nonumber \\
H_0^{\dagger} \vert j ^* \rangle = \bar{\varepsilon}_j \vert j ^* \rangle, 
\eea
and for the Floquet Hamiltonian, 
\bea
H_F \vert \lambda_{j,0} \rangle = E_{j,0} \vert \lambda_{j,0} \rangle, \nonumber \\
H_F^ {\dagger} \vert \lambda_{j,0} ^* \rangle = \overline{E}_{j,0} \vert \lambda_{j,0} ^* \rangle.
\eea
In this case the wavefunction expansion on the generalized Floquet eigenvectors is
\beq
\vert \Psi_i(t)\rangle= \sum_j e^{-i E_{j,0} t/\hbar} 
\vert \lambda_{j,0}(t)\rangle\langle \lambda_{j,0}^* (0) \vert i \rangle
\eeq
Since the Floquet Hamiltonian is non-hermitian but still symmetric (as is the case in the second application concerning
 the H$_2^+$ molecule), the normalization of the left eigenvectors can be chosen as
\beq
\vert \lambda_{j,0} ^* \rangle  = \overline{ \vert \lambda_{j,0} \rangle }.
\eeq
The wavefunction is supposed to be cyclic with respect to the subspace $S_o$ with projector $P_o = \sum_{j=1}^m |j\rangle \langle j^* |$.
 This means that the following condition is satisfied:
\begin{equation}
\forall i\leq m \;\quad  \vert \Psi_i(0) \rangle =|i\rangle \Rightarrow \vert \Psi (T) \rangle=\sum_{j=1}^m \hat{\Phi}_{ji}|j\rangle
\label{Acycl}
\end{equation}
where $\hat{\Phi}$ is now a non-unitary matrix of dimension $m$. 
Assuming that $\hat{\Phi}$ remains diagonalizable ($\hat{U}$ being the eigenvector matrix) and using the c-product normalization condition for the left eigenvectors
\cite{NHQM} ($\vert j ^* \rangle = \bar {\vert j \rangle}$), the above reasoning leads to 
a different constraint for the components of the Floquet eigenvectors: 
\begin{equation}
\sum_j e^{-\frac{i}{\hbar}E_{j,0}T -i\phi_k}\langle \tilde{k} ^*|\lambda_{j,0}(0)\rangle\langle \lambda_{j,0} ^*(0)|
\tilde{i}\rangle = \delta_{ki}, 
\label{APrinc2}
\end{equation}
with $E_{j,0}$ and $\phi_k$ complex. 
This equation is equivalent to 
\begin{eqnarray}
\sum_j d_{ji}^2 (1-e^{i\phi_{ji}})=0 \quad \mathrm{ if } \; i=k \nonumber \\
\sum_j e^{i\phi_{jk}} d_{jk} d_{ji} = 0 \quad \mathrm{ if } \; i \neq k
\label{AResbis}
\end{eqnarray}
with 
\begin{eqnarray}
\left \lbrace \begin{array}{l}
d_{ji}=\langle \lambda_{j,0}^*(0)|\tilde{i}\rangle  \; \in \mathbb{C} \\
\phi_{jk}=-\frac{1}{\hbar}E_{j,0}T-\phi_k  \; \in \mathbb{C} ,
\end{array} \right .
\label{ADefbis}
\end{eqnarray}
A solution similar to the one obtained in the hermitian case (\ref{ADiag}) still holds,
with all the $d_{ij}=0$ except for a particular value $j=j_i$ and finally
\begin{eqnarray}
\left \lbrace \begin{array}{l}
\vert \Psi_i(t) \rangle =\sum_{j=1}^m e^{-\frac{i}{\hbar}E_{j,0} t} |\lambda_{j,0}(t)\rangle \hat{U}_{ji} \\
\hat{U}_{ji}=\langle \tilde{j} ^* |i\rangle=\langle \lambda_{j,0} ^* (0)|i\rangle.
\end{array} \right .
\label{AFinbis}
\end{eqnarray}
However eq. (\ref{AResbis}) leaves open the possibility of accidental solutions with several non-zero components $d_{ij}$
and it is not possible to prove that the above solution is unique.


%
%
%
%
%
%
%
%
%
%
%
%
%
%

%

\end{document}